\documentstyle[12pt,epsf]{article}
\date  {{\small date}}     
\def\be{\begin{equation}}
\def\ee{\end{equation}}
\def\bea{\begin{eqnarray}}
\def\eea{\end{eqnarray}}
\def\beann{\begin{eqnarray*}}
\def\eeann{\end{eqnarray*}}
\newcommand{\eq}[1]{eq.~(\ref{#1})} %% to invoke write:  \eq{...}

\def\proof{ \noindent {\bf Proof:}  }

\def\limsup{\mathop{\overline{\rm lim}}}   %%%%% prettier limsup, liminf

\def\blackbox{\rightline{\vrule height 1.7ex width 1.2ex depth -.5ex}}

\newcounter{masectionnumber}
\newcommand{\masect}[1]{\setcounter{equation}{0}
  \refstepcounter{masectionnumber} \vspace{1truecm plus 1cm} \noindent
    {\large\bf \arabic{masectionnumber}. #1}\par \vspace{.2cm}
      \addcontentsline{toc}{section}{\arabic{masectionnumber}. #1}
    }

    \newcounter{masubsectionnumber}[masectionnumber]
\newtheorem{lem}{Lemma}[masectionnumber]
\newtheorem{thm}[lem]{Theorem}

\newtheorem{rem}[lem]{Remark}

\begin{document}
\title{\vspace*{-.35in}
}
\author{{\bf Oscar Bolina,}\, {\bf Pierluigi Contucci}, and
{\bf Bruno Nachtergaele}\thanks{{\bf E-mail:} bolina@math.ucdavis.edu,
contucci@math.ucdavis.edu, bxn@math.ucdavis.edu}\\[6pt]
%EndAName
Department of Mathematics\\
University of California, Davis\\
Davis, CA 95616-8633 USA\\ 
}
\title{\vspace{-1in}
{
\bf 
Path Integral Representation for Interface States of the Anisotropic
Heisenberg Model
}}
\date{}
\maketitle
\begin{abstract}
\noindent 
We develop a geometric representation for the ground state of the spin-$1/2$
quantum XXZ ferromagnetic chain in terms of suitably weighted random walks  in a
two-dimensional lattice. The path integral model so obtained admits a genuine
classical statistical mechanics interpretation with a translation invariant
Hamiltonian. This new representation is used to study the interface  ground
states of the XXZ model. We prove that the probability of having a number of
down spins in the up phase decays exponentially with the sum of their  distances
to the interface plus the square of the number of down spins. As an application
of this bound, we prove that the total third component of  the spin in a large
interval of even length centered on the interface does not fluctuate, i.e., has
zero variance. We also show how to construct a path integral representation in 
higher dimensions and obtain a reduction formula for the partition functions  in
two dimensions in terms of the partition function of the one-dimensional  model.
\newline
{\bf MCS numbers:} 82B10, 82B24, 82B41, 05A30
\end{abstract}

\masect{Introduction}
\noindent 
The advantages of a path integral representation for quantum models
have been well known since the advent of the Feynman-Kac formula.
It allows a non-commutative algebra of observables, with its hard 
algebraic problems, to be replaced by a classical configuration 
space of paths with given probability weights, thereby reducing the
computational problem to a probabilistic and combinatorial one. 
\newline
In this paper we develop a geometric representation in terms of random
paths in two dimensions for the one-dimensional spin-$1/2$ quantum XXZ 
ferromagnetic model with Hamiltonian
\be\label{HHH}
H \; = \; \sum_{x} -\frac{2}{q+q^{-1}} 
(S^{(1)}_{x} S^{(1)}_{x+1}+S^{(2)}_{x}
S^{(2)}_{x+1})-(S^{(3)}_{x} S^{(3)}_{x+1}-1/4) 
-\frac{q^{-1}-q}{2(q^{-1}+q)}(S^{(3)}_{x} - S^{(3)}_{x+1}),
\ee
where $S_{x}^{i}$ are the usual Pauli spin matrices and $0<q<1$
is a parameter that measures the anisotropy. We would like to stress,
however, that in our geometric representation the second dimension does 
{\em not\/} correspond to imaginary time, but rather to the third
component of the total spin. As in \cite{AN}, the fact that properties
related to the local spin are represented geometrically makes it
possible to derive rather strong properties about the correlations in
the ground state.

It is well-known that the model ({\ref{HHH}) has interface ground states
\cite{ASW,GW}. In the any subspace with a fixed number of down spins,
which we will call the ``canonical esemble'', the antiparallel boundary fields 
are sufficient to induce phase separation: up to order one fluctuations all 
up spins collect at one side of the interval (the left side, in the present 
case). 

In this paper we study the correlations in these interface ground states,
extending unpublished results by Koma and Nachtergaele \cite{KN}. 
Our main result is a bound on the probability of finding a number 
of down spins in the up phase at a given distance of the interface.
\vskip .05 truecm 
\noindent
{\bf Exponential bounds on the correlations.} In the canonical ensemble
in a volume $[1,N]$, with $n$ spins down, the probability of finding $v$ 
down spins located at $x_1,... ,x_v$ is bounded, uniformly in the volume,the 
by 
\be\label{UV3}
Prob(S_{x_{1}}^{z}=\downarrow, \cdots S_{x_{v}}^{z}=\downarrow) 
\leq q^{v(v-1)+2\sum_{k=1}^{v}(x_k-n)} \; ,
\ee
with $x_k-n$ being interpreted as the distance of the spin at $x_k$ to 
the interface.

This bound is similar for the ``ferromagnetic string formation probability'',
calculated for antiferromagnetic XXZ chain in \cite{EFIK}.
As an application of this bound, we prove (See Theorem \ref{thm:lim_dist})
that the  total third component of the spin in a large interval of even  length
centered on the interface does not fluctuate in the limit that the interval
tends to infinity, i.e., the distribution of this quantity tends to a Kronecker
delta. This is an a priori surprising result. A possible interpretation is
that the fluctuations of the interface can be thought of as being ``bound''
to the interface {\em and\/} occurring in pairs, similar to particle-hole pairs.

\noindent
The paper is organized as follows. In {\it Section 2} we introduce 
path integral models for weighted random walk in two dimensions.
In {\it Section 3} we show how to relate the ground state property 
of the quantum model to the correlation functions of a suitable weighted
random walk. A classical statistical mechanics interpretation of the 
path integral model is introduced in {\it Section 4}. In {\it Section 5}
we prove a Markov-type property for the partition functions and also
the action of the translation group. In {\it Sections 6} and {\it 7} 
we prove the bound (\ref{UV3}) and apply it to the fluctuations of the third
component of the spin. In {\it Section 8} we consider
higher dimensional models and prove a dimensional reduction formula 
for the partition functions in two-dimensions in terms of the partition
functions of the one-dimensional model.

\masect{Path Integral Models in the Two-Dimensional Lattice.}
\noindent
Let ${\bf Z}^2_{+}$ be the set of points in the positive quadrant 
of the two dimensional lattice ${\bf Z}^2$. A ``zig-zag'' path from 
the origin $(0,0)$ to some final point $(n,m)$ is a connected path 
in ${\bf Z}^2_{+}$ monotonically increasing in both coordinates.
Its length (the sum of the steps) is equal to 
$L=n+m$, as shown in Fig. 1.
\begin{figure}
\centerline{\epsfbox{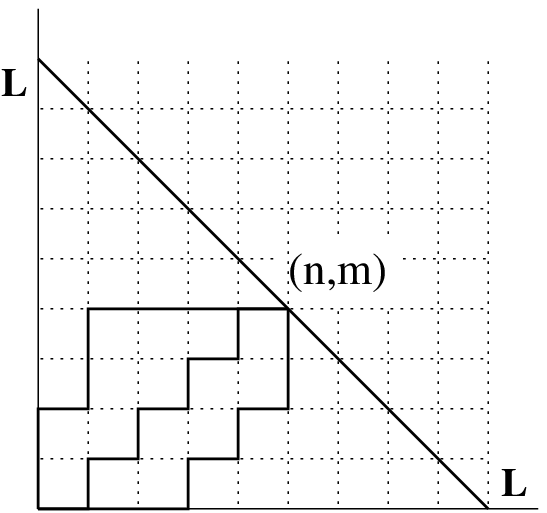}}
\caption{{\sl Three paths on ${\bf Z}^{2}_{+}$ from the origin to
$(n,m)$}}
\end{figure}
\noindent
A path integral model on ${\bf Z}^2_{+}$ is a law that associates
positive weights $w(p)$, to each path $p$ in the lattice.
\newline
%A general vector in this Hilbert space is a linear combination
%\be
%\mid \psi \rangle = \sum_{p \in {\cal P}_{(n,m)}} c(p) \mid p \rangle
%\ee
%where ${\cal P}_{(n,m)}$ is the set of all paths from the origin to
%$(n,m)$.
%\newline
%The norm of a such a vector is given by
%\be\label{N}
%Z(n,m)=\sum_{p \in {\cal P}_{(n,m)}} w^{2}(p).
%\ee
We denote by ${\cal P}_{(n,m)}$ the set of all paths from the origin to
a point $(n,m)$ and define the canonical partition function
\be
Z(n,m)=\sum_{p \in {\cal P}_{(n,m)}} w(p) \; 
\label{N}
\ee
This formalism can be extended to ``zig-zag'' paths which go from 
any arbitrary origin $(n',m')$ to the final point $(n,m)$ with 
$n'\le n$ and $m' \le m$. We call this set of paths 
${\cal P}_{(n',m'; n,m)}$, and define a {\it generalized} 
partition function by
\be\label{GN}
Z(n',m';n,m)=\sum_{p \in {\cal P}_{(n',m';n,m)}} w(p).
\ee
In path integral models, correlation functions measure the probability
that a path goes through particular points $(x_1,y_1), (x_{2},y_{2}), 
\cdots, (x_{r},y_{r})$. The one-point correlation function is defined
as the probability of crossing the point $(x,y)$
\be\label{prob1}
P_{n,m}(x,y)=\frac{Z(n,m \mid x,y)}{Z(n,m)},
\ee
where
\be
Z(n,m \mid x,y) \; = \; \sum_{p\in {\cal P}_{(n,m)}(x,y)} w(p)
\label{condpar}
\ee
and ${\cal P}_{(n,m)}(x,y)$ is the set of paths from the origin to
$(n,m)$ that pass through the point $(x,y)$. More generally, we can 
define
\be\label{probr}
P_{n,m}(x_1,y_1;\cdots;x_r,y_r)=
\frac{Z(n,m \mid x_1,y_1; \cdots; x_r,y_r)}{Z(n,m)},
\ee
where
\be
Z(n,m \mid x_1,y_1; \cdots; x_r,y_r)
\; = \; \sum_{p\in {\cal P}_{(n,m)}(x_1,y_1; \cdots; x_r,y_r)} w(p)
\label{condparr}
\ee
and ${\cal P}_{(n,m)}(x_1,y_1; \cdots; x_r,y_r)$ denotes the set of 
paths that pass through the particular points $(x_1,y_1), (x_{2},y_{2}), 
\cdots, (x_{r},y_{r})$. 
\newline
In this framework, we consider models for which the weight $w(p)$
is a local function of the bonds that the path is passing through. 
Denoting by ${\bf B}^{2}_{+}$ the set of bonds in ${\bf Z}^2_{+}$, 
we associate a positive number $w(b)$ to each element $b$ of 
${\bf B}^{2}_{+}$ and define
\be
w(p)=\prod_{b \in p} w(b).
\ee
This formalism admits a generalization when, instead of restricting
the paths to reach one final point, we extended it to all paths of 
given length $L=n+m$ (the grand-canonical ensemble). In this way we 
define the grand-canonical partition function
\be
\tilde{Z}(L)=\sum_{p \in {\cal P}_{L}} \tilde{w}(p).
\label{N1}
\ee
where ${\cal P}_L=\cup_{n+m=L}{\cal P}_{n,m}$.
\newline
The relation between the partition functions (\ref{N}) and 
(\ref{N1}) is made particularly useful when we chose 
$\tilde{w}(p)=z^n w(p)$ where $n$ is the horizontal displacement 
of $p$. In this case we get the following generating function 
relation
\be
\tilde{Z}(L)(z)=\sum_{n=0}^{l}z^n Z(n,L-n)
\label{N2}
\ee

\masect{The One-Dimensional Spin-$1/2$ XXZ Ferromagnetic Model}
\noindent
The path integral formalism developed in the previous section provides a
geometric representation for interface ground state of quantum spin 
systems governed by the XXZ Hamiltonian.
\newline
In one dimension, the Hamiltonian for the spin-$1/2$ XXZ ferromagnetic
chain of length {\it L} with special boundary terms is given by 
\cite{ASW,GW}
\be\label{H}
H_{L}=\sum_{x=1}^{L-1} h_{x,x+1},
\ee  
where 
\be
h_{x,x+1}=-\Delta^{-1} (S^{(1)}_{x} S^{(1)}_{x+1}+S^{(2)}_{x}
S^{(2)}_{x+1})-(S^{(3)}_{x} S^{(3)}_{x+1}-1/4) 
-A(\Delta)(S^{(3)}_{x} - S^{(3)}_{x+1}).
\ee
Here $S^{i}_{x}$ ($i=1,2,3$) are the usual Pauli spin matrices at
the site {\it x}, $\Delta \geq 1$ is the anisotropy parameter and
$A(\Delta)$ is a boundary magnetic field given by
\be
A(\Delta)=\frac{1}{2}\sqrt{1-\Delta^{-2}}. 
\ee
A configuration of spins in the one dimensional chain is identified
with the set of numbers ${\alpha_x}$ for $x=\{1,2,..., L\}$
where $\alpha$ takes values in the set $\{0,1\}$. We choose $\alpha=0$
to correspond to an up spin, or, in the particle language, to an
unoccupied site. Conversely, $\alpha=1$ corresponds to a down spin or 
an occupied site. It can be proved \cite{ASW,GW} that the ground state 
of the model in the sector with {\it n} down spins is given by 
\be\label{VN}
\psi(n,m) =\sum_{\{\alpha_{x}\}\in {\cal A}_{n,m} } 
\left \{ \prod_{x=1}^{L} q^{\alpha_{x}x} \right \} \mid \{ \alpha_{x} \}
\rangle
\ee
where the ${\cal A}_{n,m}$ the set of configurations $\{ \alpha_{x} \}$
such that $\sum_{x} \alpha_{x}=n$, and the real and positive parameter 
{\it q} is defined in term of the anisotropic coupling by 
\be
\Delta =\frac{q+q^{-1}}{2} \;\;\; {\rm with} \;\;\; 0 < q < 1.
\ee
The norm of the ground state vector (\ref{VN}) with {\it n} spins down is
\be\label{norm}
\| \psi(n,m) \|^2=\sum_{\{\alpha_{x}\}}~
\prod_{x=1}^{L}~ q^{2x \alpha_{x}} 
\ee
\newline 
To construct the classical path integral representation for the
quantum {\it XXZ} model, we identify the norm (\ref{norm}) of 
the ground state vector (\ref{VN}) with the canonical partition 
function (\ref{N}) in the path integral formalism by assigning 
suitable weights to the bonds of the corresponding two dimensional 
path space.

\begin{thm}[Path integral representation for interface ground state]
\be\label{huh}
\| \psi(n,m) \|^2=:Z(n,m)=\sum_{p \in {\cal P}_{(n,m)}} w(p) \; 
\label{N11}
\ee
is the partition function for the classical path integral 
model associated with the quantum {\it XXZ} model for the 
the following choice of weights
\be\label{bond1}
w(b)=\left \{
\begin{array}{ll}
q^{2(x_b+y_b)} \;\;\; 
{\rm for~ a~ horizontal~ bond~ whose~ right~ end~ is~ at}~ (x_b,y_b) \\
1 \;\;\;\;\;\;\;\;\;\;\;\;\;\; {\rm any~ vertical~ bond} \; .
\end{array}
\right.
\ee
\end{thm}
\proof
From expression (\ref{norm}) we have
\be
\sum_{\{\alpha_{x}\}}~
\prod_{x=1}^{L}~ q^{x \alpha_{x}} 
= \sum_{1 \leq x_{1} < x_{2} < ... < x_{n} \leq L}
q^{2(x_{1}+...+x_{n})} \; ,
\label{rrepp}
\ee
where the $x_{i}$ are the positions of the down spins in the chain.
Observing that the position of a down spin in the lattice is equal to 
the distance of a given point in the path from the origin 
$x_i = x_b+y_b$, \eq{bond1} follows. 
\newline 
\blackbox

\masect{Classical Statistical Mechanics Interpretation}
\noindent
The paths integral models treated so far admit a classical
statistical mechanics interpretation, based on the following
result
\begin{thm}\label{areath}
Given an element of ${\cal P}_{n,m}$ we define the {\it area}
of a path by (see Fig. 2)
\be
A(p) \; = \; \# \{\rm plaquettes~ under~ {\sl p} \}.
\label{defa}
\ee
We have
\be
w(p)=n(n+1)+2A(p) \; .
\label{area}
\ee
\end{thm}
{\bf Proof.} The theorem is true, by inspection, for the path of 
minimum weight, which is the path $\tilde{p}$ that goes through
$(n,0)$. In this case 
\be
w(\tilde{p})=2 \sum_{j=1}^{n} j.
\label{wju}
\ee
Any other path can be obtained from the minimum weight path by 
the application of a local operation {\it C} that adds a plaquette 
to a concave corner in such a way that the weight of the path
path obtained is
\be
w(Cp) \; = \; 2 + w(p) \; 
\label{cip}
\ee
in accordance with \eq{bond1}.
\newline 
\blackbox
\begin{rem}
The area of a path can be regarded as the  Hamiltonian of a corresponding
classical statistical model with  the partition function
\be\label{pfa}
Z(n,m)=q^{n(n+1)} \sum_{p \in {\cal P}_{(n,m)}} e^{-\beta H(p)}.
\ee
with the identification $H(p)=A(p)$ and $q^2=e^{-\beta}$, for $0 < q < 1$.
\end{rem}
The former property allows us to prove the main result of this section.
\begin{figure}
\centerline{
\epsfbox{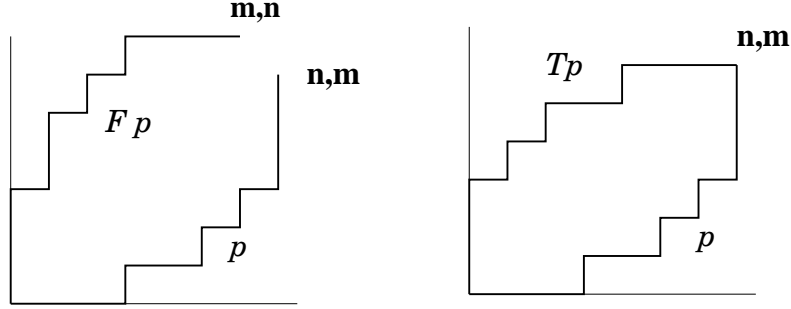}}
\caption{{\sl Parity and time reversal symmetries}}
\end{figure}
\noindent
\begin{thm}\label{symm}
Consider the following transformations in the space of paths ${\cal P}_L$: 
\newline
{\it 1}. {\rm The parity}
\be
F: \; p\in{\cal P}_{n,m} \to F(p)\in{\cal P}_{m,n} \; ,
\label{pi}
\ee
is defined by the reflection with respect to diagonal (see Fig.2 ). If 
$p$ corresponds to the sequence $\alpha_1,\alpha_2,\cdots, \alpha_{L}$,
$F(p)$ corresponds to $1-\alpha_1,1-\alpha_2,\cdots,1-\alpha_{L}$.
\newline
{\it 2.} {\rm The time reversal}
\be
T: \; p\in {\cal P}_{n,m} \to T(p)\in{\cal P}_{n,m} \; ,
\label{ti}
\ee
is defined by the time reversed path (Fig. 2). If $p$ is 
$\alpha_1,\alpha_2,\cdots, \alpha_{L}$, $T(p)$ is 
$\alpha_{L}, \alpha_{L-1}, \cdots, \alpha_1$.
\newline
The combined transformation is a symmetry for our path integral model
in the sense that
\be
Prob(p) \; = \; Prob(FT(p)) \; 
\label{areat}
\ee
\end{thm}
{\bf Proof:} We clearly have
\be
A(p)+A(Fp)=nm,
\label{pa}
\ee
and
\be
A(p)+A(Tp)=nm.
\label{tr}
\ee
By applying \eq{pa} to a time reversed path $T(p)$ we see that 
$A(Tp)+A(TFp)=nm$. This, together with (\ref{tr}) leads to 
$A(p)=A(FTp)$ 
\newline 
\blackbox
\begin{lem}\label{lem:4.3}
The partition function $Z(n,m)$ for paths ${\cal P}_{n,m}$ and the
partition function $Z(m,n)$ for time reversed paths ${\cal P}_{m,n}$ 
satisfy the relation
\be
\frac{Z(n,m)}{q^{n(n+1)}}=\frac{Z(m,n)}{q^{m(m+1)}}.
\label{efhv}
\ee
\end{lem}
\proof The result is a consequence of (\ref{pfa}) and the fact that 
both the transformations $F$ and $T$ are one to one. 
\newline 
\blackbox
\newline
With the aid of Theorem \ref{thm:translations}, proved in the next section,
this property can be extended to the generalized partition functions. 
\begin{lem}
\be
\frac{Z(n',m';n,m)}{q^{(n+m')(n+m'+1)}}=\frac{Z(m',n';m,n)}
{q^{(n'+m)(n'+m+1)}}
\ee
\end{lem}
\proof We first shift the generalized partition function to the 
the origin with the translation property formula (\ref{TF}). This gives 
\be\label{a}
Z(n',m';n,m)=q^{2(n'+m')(n-n')}~Z(n-n',m-m')
\ee
Next we use (\ref{lem:4.3}) to rewrite $Z(n-n',m-m')$ above in 
terms of $Z(m-m',n-n')$. We obtain
\be\label{b}
Z(n',m';n,m)=q^{(n-n')(n+n'+2m'+1)-(m-m')(m-m'+1)}Z(m-m',n-n')
\ee
Now we again use the translation property to shift $Z(m-m',n-n')$ 
back to $Z(m',n';m,n)$ and get
\be\label{c}
Z(m',n';m,n)=q^{2(n'+m')(m-m')}~Z(m-m';n-n').
\ee 
The lemma follows from (\ref{c}) and (\ref{b}).
\newline
\blackbox

\masect{Geometric properties of Z}
\noindent
In this section we study the properties of the partition function 
(\ref{huh}) and the corresponding generalized partition function 
(\ref{GN}) associated with the {\it XXZ} model.
\newline
The two main properties we prove are a {\it Markov type} property 
and the action of the translation group on partition functions. 
This two properties together provide two independent relations 
that {\it solve explicitly} the one-dimensional quantum system.
\newline
We have the following theorem. 
\begin{thm}[Markov property]\label{t1}
For any integer $z$ such that $n'+m' \le z \le n+m$ 
\be
Z(n',m';n,m) \; = \; \sum_{x+y=z} Z(n',m';x,y)Z(x,y;n,m)
\label{genrec}
\ee
See Fig. 3. for a pictorial representation.
\end{thm}
\begin{figure}
\centerline{
\epsfbox{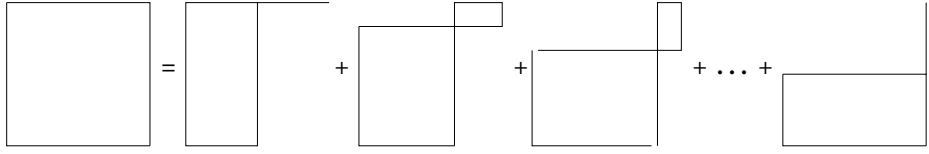}}
\caption{{\sl Graphical representation of the Markov property (\ref{t1})
of the partition function}}
\end{figure}
\noindent
\proof We write (\ref{GN}) with the set of path  
${\cal P}_{(n',m';n,m)}=\cup_{x+y=z} {\cal P}_{(n',m';n,m)}(x,y)$
as
\be\label{GNN}
Z(n',m';n,m)=\sum_{ \cup_{x+y=z} {\cal P}_{(n',m';n,m)}(x,y)} w(p). 
\ee
Replacing the sum over the union of paths with an extra sum over the 
paths, we get
\begin{eqnarray}\label{GNN2}
Z(n',m';n,m)&=&\sum_{x+y=z} \sum_{p \in {\cal P}_{(n',m';n,m)}(x,y)} 
w(p), \nonumber\\
&=& \sum_{x+y=z} Z(n',m';x,y)Z(x,y;n,m) \; ,
\end{eqnarray}
where the last equality comes from the fact that our paths are 
monotonically increasing.
\newline
\blackbox
\newline
In the particular case we restrict the sum over {\it z} in
{\it theorem 5.1} to be over two points for which $z=n+m-1$,
the partition function $Z(n,m)$ satisfies the recursion 
relation (see Fig. 4) given in the following lemma.
\begin{lem}
\be\label{uppercorner}
Z(n,m)=Z(n,m-1)+q^{2(n+m)}Z(n-1,m).
\ee
\end{lem}
\proof Follows from {\it Theorem 5.1} with the weights (\ref{bond1}).
\newline
\blackbox
\newline
Formula (\ref{uppercorner}) relates the two nearest neighbors of the
final point $(n,m)$ in the upper right corner of Fig. 2. A similar
relation can be devised between the two nearest neighbors of the
initial point $(0,0)$ in the lower left corner. We have
\begin{lem} The partition function $Z(n,m)$ satisfies the following 
recursion relation in terms of generalized partition functions (see 
Fig. 5) 
\be\label{lowercorner}
Z(n,m)=q^{2}Z(1,0;n,m)+Z(0,1;n,m).
\ee
\end{lem}
\proof Follows from the same reasoning that led to (\ref{uppercorner}).
\newline
\blackbox
\newline
\begin{figure}
\centerline{
\epsfbox{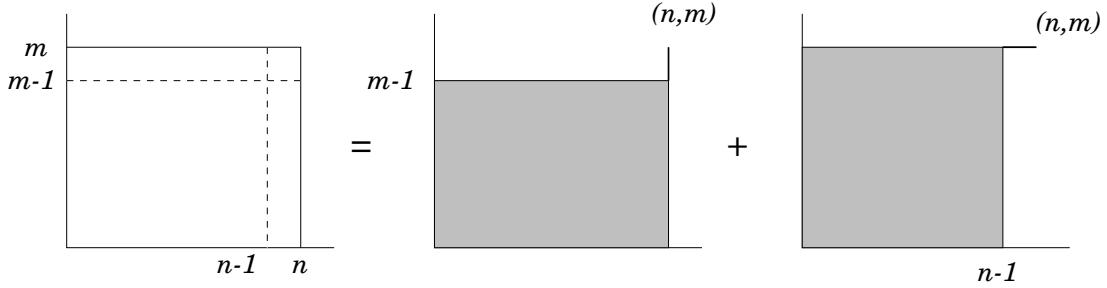}}
\caption{{\sl Graphical representation of the recursion relation
(\ref{uppercorner})}}
\end{figure}
\noindent
Note that (\ref{lowercorner}), unlike (\ref{uppercorner}),
involves generalized partition functions. However, the action 
of the translation group on the generalized partition
function in (\ref{lowercorner}) transforms them in ordinary
partition functions by means of multiplication factor. We have
\begin{thm}[Action of the translation group]\label{thm:translations}
For every $x\le n'$ and $y\le m'$
\be\label{TF}
Z(n',m'; n,m)=q^{2(x+y)(n-n')} Z(n'-x,m'-y;n-x,m-y),
\ee
\end{thm}
\proof 
We first note that $Z(n',m';;n,m)$ is a polynomial in {\it q} that
can be written as
\be\label{P1}
Z(n',m';n,m)=q^{r}(1+a_{1}q^{2}+a_{2}q^{4}+ ... + a_{(m-m')(n-n')}
q^{2(m-m')(n-n')})
\ee
where $r=[2(m'+1)+2n'](n-n')+(n-n')(n-n'-1)$ is the minimum power
of {\it q} among all the paths from $(n',m')$ to $(n,m)$, and the
(positive) coefficients $a_{j}$ account for the multiplicity
of the powers of $q^{2j}$. Namely given the box $B(n',m';n,m)$:
\be
a_{j} \; = \; \# \{{\rm paths~ in}~B(n',m';n,m)~ |~ A(p)=j \}.
\label{multi}
\ee
If we perform a shift {\it x} in the horizontal direction and a shift
{\it y} in the vertical direction, we obtain the translated partition
function
\be\label{P2}
Z(n'-x,m'-y;n-x,m-y)=q^{r'}(1+a_{1}q^{2}+a_{2}q^{4}+ ... +
a_{(m-m')(n-n')} q^{2(m-m')(n-n')})
\ee 
where $r'=[2(m'-y+1)+2(n'-x)](n-n')+(n-n')(n-n'-1)$, and the polynomial 
inside the parenthesis on the right hand side of (\ref{P2}) is the 
{\it same} as in (\ref{P1}) because of the translation invariance of
the {\it area} Hamiltonian in \eq{pfa}.
Consequently 
\be
Z(n',m';n,m)=q^{(r-r')}Z(n'-x,m'-y;n-x,m-y),
\ee
which is just (\ref{TF}). 
\newline
\blackbox
\newline
The application of the translation property (\ref{TF}) to the
partition function in (\ref{lowercorner}) provides a second 
independent relation between the partition functions containing 
only the nearest neighbors of the point $(n,m)$. 
\begin{lem} The partition function satisfies 
\be\label{RR3}
Z(n,m)=q^{2n}Z(n-1,m)+q^{2n}Z(n,m-1).
\ee
\end{lem}
\begin{figure}
\centerline{
\epsfbox{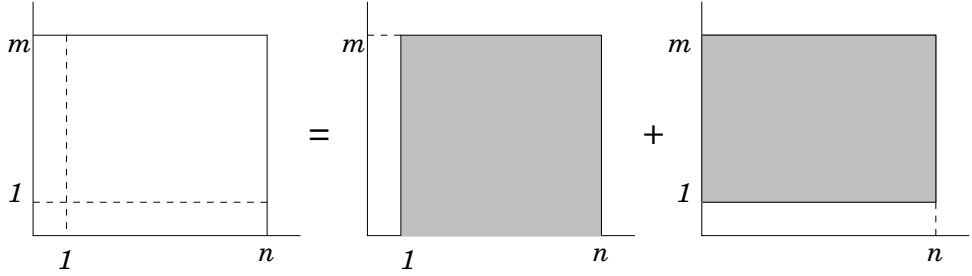}}
\caption{{\sl Graphical representation of the recursion relation
(\ref{lowercorner})}}
\end{figure}
\noindent
\proof Direct application of the translation property of (\ref{TF})
allows us to write the generalized partition functions in 
(\ref{lowercorner}) in terms of ordinary partition functions
\be \label{RR2}
Z(1,0;n,m)=q^{2(n-1)}Z(n-1,m) \;\;\;\; {\rm and} \;\;\;\;
Z(0,1;n,m)=q^{2n} Z(n,m-1)
\ee
Substituting (\ref{RR2}) in (\ref{lowercorner}) yields the lemma. 
\newline
$\blackbox$
\begin{rem}
It is important to emphasize that the path integral formalism generated two {\it
independent} relations between $Z(n,m)$,  $Z(n-1,m)$ and $Z(n,m-1)$, namely
(\ref{uppercorner}) and  (\ref{RR3}), which are known as the q-Pascal identities
for Gauss polynomials \cite{Kas}. This fact is reminiscent of the situation
found in the general theory of stochastic processes in which conditioning the
process with respect to the initial  or the final conditions provides two
independent relations.
\end{rem}
The independence of the two relations 
allow us to derive an 
explicitly expression for the partition function (\ref{P1}) 
as a product formula. 
\begin{thm} The partition function $Z(n,m)$ is given by 
\be
Z(n,m)=q^{n(n+1)}~ \frac{\prod_{i=1}^{n+m} (1-q^{2i})}{
\prod_{i=1}^{n} (1-q^{2i})~ \prod_{i=1}^{m} (1-q^{2i})}.
\ee
\end{thm}
\proof Solving (\ref{uppercorner}) and (\ref{lowercorner}) for $Z(n-1,m)$
and $Z(n,m-1)$ in terms of $Z(n,m)$ we get
\be\label{rel}
\frac{Z(n-1,m)}{Z(n,m)}=q^{-2n} \frac{1-q^{2n}}{1-q^{2(n+m)}} \;\;\;\;
{\rm and} \;\;\;\; \frac{Z(n,m-1)}{Z(n,m)}= \frac{1-q^{2m}}{1-q^{2(n+m)}}.
\ee
From (\ref{rel}) we obtain  
\be\label{ic}
\frac{Z(n-1,m-1)}{Z(n,m)}= q^{-2n}~ \frac{(1-q^{2n})(1-q^{2m})}
{(1-q^{2(L-1)})(1-q^{2L})}
\ee
Setting the initial condition $Z(0,0)=1$ yields the theorem.
\newline
\blackbox

\begin{lem} For $v \leq n$, $w \leq m$, the partition 
function $Z(n-v,m-w)$ satisfies
\be\label{wow6}
Z(n-v, m-w) \leq q^{-2nv+v(v-1)}~Z(n, m).
\ee
\end{lem}
\proof Starting from (\ref{ic}) and successively applying the 
first of the recursion relations (\ref{rel}) {\it v} times, we 
obtain 
\be\label{wtw1}
Z(n-v,m-1)=K_{v-1} L_{v-2} ...  L_{0} Z(n,m)
\ee 
where, according to (\ref{rel}) and (\ref{ic}), we define
\[
K_{j}=q^{-2(n-j)}~ \frac{(1-q^{2(n-j)})(1-q^{2m})}
{(1-q^{2(n-j+m-1)})(1-q^{2(n-j+m)})} \;\;\;\; {\rm and} \;\;\;\;
L_{j}=q^{-2(n-j)} \frac{1-q^{2(n-j)}}{1-q^{2(n-j+m)}}.
\]
Now, successively applying the second of the recursion relations
(\ref{rel}) {\it w} times, we obtain
\be\label{wtw2}
Z(n-v,m-w)=M_{w-1}... M_{1}Z(n,m)
\ee
where
\[
M_{j}=\frac{1-q^{2(m-j)}}{1-q^{2(n+m-j)}}
\]
Combining (\ref{wtw1}) and (\ref{wtw2}) gives
\be
\frac{Z(n-v,m-w)}{Z(n,m)} \leq \prod_{j=0}^{v-1} q^{-2(n-j)}
\ee
and the theorem follows.
\newline
\blackbox
\newline

\masect{Probability Estimates}
\noindent
In this section we show how to bound the correlation functions
for the quantum model through bounds on the path integral model
correlations functions. The probability that a given spin, or
a set of spins, are up or down can be expressed as sums of probabilities
that a path goes through a given, or many, bonds.
\newline
The path integral representation provides a remarkable pictorial
interpretation of these probabilities which allows us to obtain 
the estimates in an elementary way by efficiently exploiting the 
action of the translation group.
\newline
Our first result is the
\begin{thm} The probability that a path from the origin to $(n,m)$
pass through the point $(x,y)$ is given by
\be
P_{n,m}(x,y)=q^{2(x+y)(n-x)} \frac{Z(x,y)Z(n-x,m-y)}{Z(n,m)}.
\ee
\end{thm}
\proof By the one-point correlation function (\ref{prob1})
we have
\be\label{prb}
P_{n,m}(x,y)=\frac{Z(n,m \mid x,y)}{Z(n,m)},
\ee 
where $Z(n,m \mid x,y)$ is the number of paths from the origin to 
$(n,m)$ passing through the point $(x,y)$. By {\it Theorem 4.1} we 
also have 
\be\label{prob2}
Z(n,m \mid x,y)=Z(x,y) Z(x,y; n,m).
\ee
Now we use the translation property (\ref{TF}) to shift
$Z(x,y;n,m)$. We obtain
\be\label{shift}
Z(x,y;n,m)=q^{2(x+y)(n-x)} Z(n-x,m-y).
\ee
Substituting (\ref{shift}) in (\ref{prb}) yields the Theorem.
\newline
\blackbox
\newline
As to the probability that a path goes through a particular bond,
we have the following estimates (which is useful for $x \geq n$)
\begin{thm}
Considering the quantity
\be
P(S^{z}_{x}=-1):=\frac{\langle \psi(n,m) \mid (1/2-S^{z}_{x})~
\psi(n,m) \rangle}{ \| \psi(n,m) \|^{2}} \; ,
\label{defn}
\ee
we have that
\be
P_{n,m}(S^{z}_{x}=-1) \; = \; \sum_{j=x-m}^{n}P_{n,m}(j-1,x-j;j,x-j)
\label{pi=qm} \; 
\ee
and the following bound holds
\be\label{hgf}
P_{n,m}(S^{z}_{x}=-1) \leq q^{2(x-n)} \frac{1-q^{2n}}{1-q^{2(n+m)}}.
\ee
\end{thm}
\begin{figure}
\centerline{
\epsfbox{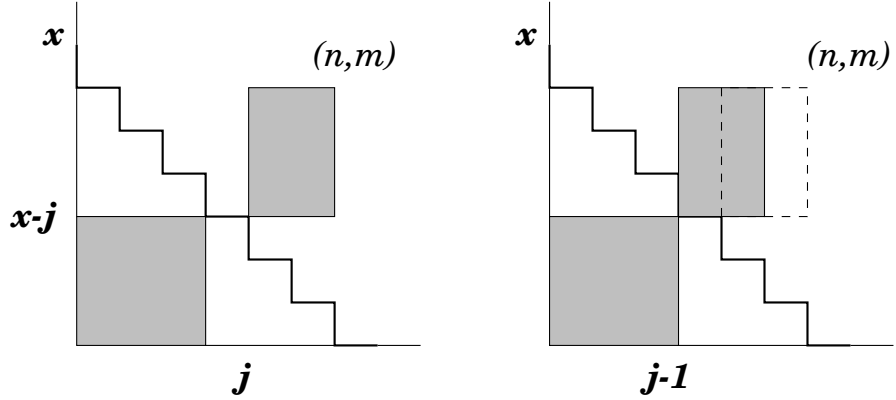}}
\caption{{\sl The paths in which the $x^{th}$ spin is down are contained
in the two shaded areas in the diagram. To obtain the probability
(\ref{prob0}) we shift the upper box $B(j,x-j;n,m)$ one unit to the 
left and sum along the line {\bf x}.}}
\end{figure}
\noindent
\begin{rem}
We have seen in {\it Section 4} that the one-dimensional XXZ model 
and its ground states are invariant under the combined spin flip and
left-right symmetries. This fact is implies the property
\be\label{ijk}
P_{n,m}(S_{x}^{z}=-1)=P_{m,n}(S_{L-x+1}^{z}=+1) \; ,
\ee  
In this way the properties that we are proving for $x\ge n$ can 
be transformed in the similar ones for $x\le n$.
\end{rem}
\proof To obtain the probability that the $x^{th}$ spins is down, we 
have to sum the probabilities that the paths from $(0,0)$ to $(n,m)$ 
go {\it horizontally} through the diagonal line in which the sum of 
the coordinates is {\it x} because each horizontal bond in the path
represent a down spin.
\newline
The graphical representation of this probability is shown in Fig 6
for $x \geq n$, $x \geq m$, and $x < n+m$. In this case, the $x^{th}$ 
step has to be taken in the horizontal direction. Then we have  
\be\label{prob0}
P_{n,m}(S_{x}=-1)=\sum_{j=x-m}^{n} P_{n,m}(j-1,x-j;j,x-j)
\ee
where $P_{n,m}(j-1,x-j;j,x-j)$ is the probability that the path goes
through the bond $(j-1,x-j) \rightarrow (j,x-j)$ is given by
\be\label{yy}
P_{n,m}(j-1,x-j;j,x-j)
=q^{2x} \frac{Z(j-1,x-j) Z(j,x-j;n,m)}{Z(n,m)}.
\ee
We see that each $P_{n,m}(j-1,x-j;j,x-j)$ is represented by a box from
the origin to the tip of a horizontal bond on the sphere of radius {\it
x}, connected to another box from the tip of the horizontal bond to the
final point $(n,m)$. 
\newline
A bound on $P_{n,m}(j-1,x-j;j,x-j)$ is the result of an operation we
perform on Fig. 6, by shifting the upper box in the figure one unit to
the left in the horizontal direction, as indicated. This is the same as
making  an equal shift on $Z(j,x-j;n,m)$. By the translation property
(\ref{TF}), we have
\be
Z(j,x-j; n,m)=q^{2(n-j)} Z(j-1,x-j;n-1,m).
\ee
We thus get
\be\label{prob4}
P_{n,m}(S_{x}=-1)=q^{2x}~\sum_{j=x-m}^{n} \frac{Z(j-1,x-j)
Z(j-1,x-j;n-1,m)}{Z(n,m)}~ q^{2(n-j)}.
\ee
The easy bound follows immediately
\be\label{prob7}
P_{n,m}(S_{x}=-1) \leq q^{2x}~\sum_{j=1}^{x} \frac{Z(j-1,x-j)
Z(j-1,x-j;n-1,m)}{Z(n,m)},
\ee
and, by {\it Theorem 4.1}, the summation over {\it j} is nothing 
more than the partition function $Z(n-1,m)$.
\newline
Thus we get
\be\label{prob8}
P_{n,m}(S_{x}=-1) \leq q^{2x}~\frac{Z(n-1,m)}{Z(n,m)}.
\ee
We have worked out the ratio ${Z(n-1,m)}/{Z(n,m)}$ in {\it Section 4}.
The substituting of formula (\ref{rel}) of that section in (\ref{prob8})
gives the theorem. 
\newline
\blackbox
\newline
The same reasoning with minor changes leads to the estimates for 
the probability that the $x^{th}$ spin is up. We have 
\begin{thm}
\be
P_{n,m}(S^{z}_{x}=+1) \leq \frac{1-q^{2m}}{1-q^{2(n+m)}}.
\ee
\end{thm}
\proof As depicted in Fig.7, also for $x \geq n$, $x \geq m$ and
$x \leq n+m$, we have
\be\label{prob9}
P_{n,m}(S_{x}=+1)=\sum_{j=x-m}^{n} P_{n,m}(j,x-j-1;j,x-j)
\ee
where $P_{n,m}(j,x-j-1;j,x-j)$ is the probability that the path goes 
through the bond $(j,x-j-1) \rightarrow (j,x-j)$ is given by
\be
P_{n,m}(j,x-j-1;j,x-j)=\frac{Z(j,x-j-1) Z(j,x-j;n,m)}{Z(n,m)}.
\ee
By applying the translation property (\ref{TF}) to 
$Z(j,x-j;n,m)$ to shift it one unit down in the vertical 
direction, we get
\be
Z(j,x-j; n,m)=q^{2(n-j)} Z(j,x-j-1;n,m-1).
\ee
Substituting the above relation in (\ref{prob9}) gives
\be\label{prob10}
P_{n,m}(S_{x}=+1) \leq \sum_{j=x-m}^{n}
\frac{Z(j,x-j-1)Z(j,x-j-1;n,m-1)}{Z(n,m)} = \frac{Z(n,m-1)}{Z(m,n)}.
\ee
By inserting (\ref{rel}) into the above expression gives the
theorem. 
\newline
\blackbox
\begin{figure}
\centerline{
\epsfbox{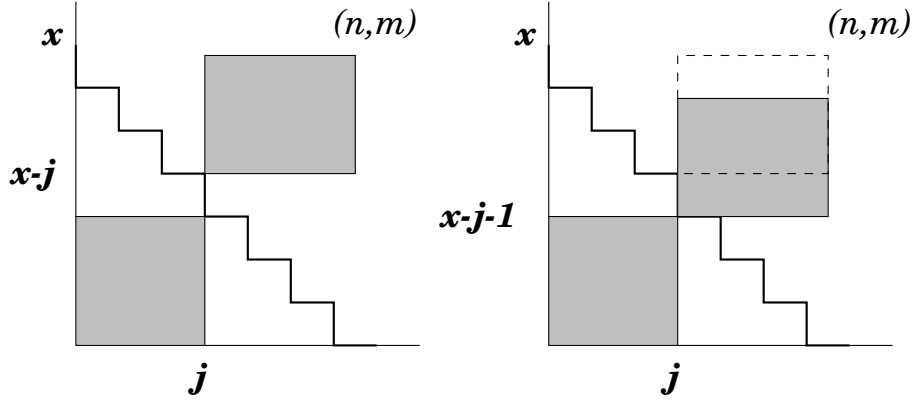}}
\caption{{\sl To obtain the probability (\ref{prob9}) we shift the upper 
box down one unit and sum along the line {\bf x}.}}
\end{figure}
\noindent
Finally, we have to consider the probability that adjacent spins are
opposite. We suppose that the $x^{th}$ spins is down and the 
$(x+1)^{th}$ spins is up. We prove that
\begin{thm}
\be\label{+-}
P_{n,m}(S_{x}^{z}=-1,S_{x+1}^{z}=+1) \leq q^{2(x-n)}
\frac{1-q^{2m}}{1-q^{2n}} \frac{1-q^{2L}}{1-q^{2(L-1)}}
\ee
\end{thm}
\proof For $x \geq n$, $x \geq m$ and $x < n+m$, we have
\be\label{kk}
P_{n,m}(S_{x}^{z}=-1,S_{x+1}^{z}=+1)=\frac{q^{2x}}{Z(n,m)}
\sum_{j=x-m}^{n} Z(j-1,x-j)~Z(j,x-j+1;n,m)
\ee 
By performing a translation along both the horizontal and vertical 
direction by one unit, as in Fig. 8, we bring the origin of $Z(j,x-j+1;
n,m)$ to the point $(j-1,x-j)$, thus obtaining
\be
Z(j,x-j+1;n,m)=q^{4(n-j)}Z(j-1,x-j;n-1,m-1).
\ee
Substitution in (\ref{kk}) yields
\begin{eqnarray}\label{kk1}
P_{n,m}(S_{x}^{z}=-1,S_{x+1}^{z}=+1) &\leq& q^{2x} \sum_{j=x-m}^{n}
\frac{Z(j-1,x-j)Z(j-1, x-j;n-1,m-1)}{Z(n,m)} \nonumber \\
&=& q^{2x} \frac{Z(n-1,m-1)}{Z(n,m)}
\end{eqnarray}
and the theorem follows from (\ref{ic}). 
\begin{figure}
\centerline{
\epsfbox{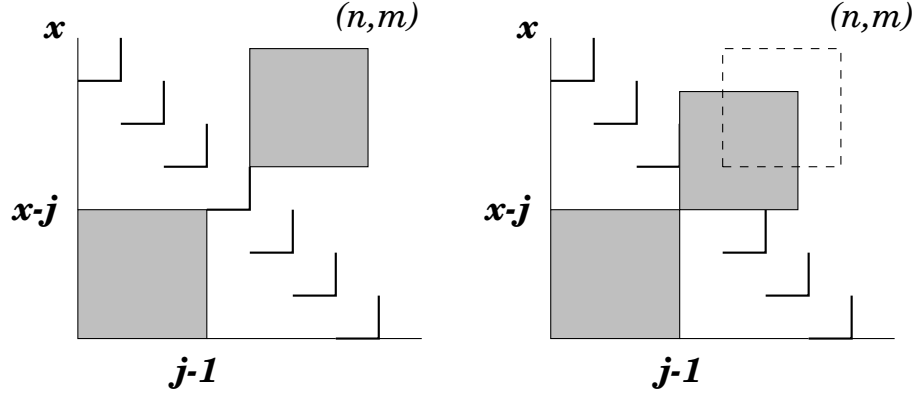}}
\caption{{\sl To obtain the probability (\ref{kk}) we shift the upper 
box to the left and down by one unit and sum along the line {\bf x}.}}
\end{figure}
\noindent

\masect{Multi-Point Correlation Functions}
\noindent
We now extend the analysis of the previous section to 
include the probability that a string of {\it r} spins 
at positions $x_1, x_2, \cdots, x_r$ has a given 
configuration of up or down spins. 
\newline
Let us consider multi-point correlation functions as given in the
definitions (\ref{probr}) and (\ref{condparr}), and paths from the 
origin to $(n,m)$ that cross successive spheres on the lattice of 
radius $x_1, x_2 \cdots, x_r$. We take the case in which $x_{j}> 
n,m$ and $x_{j} \leq n+m$ for $j=1,2, \cdots, r$.  
\newline
We denote by $P_{n,m}(S_{x_{1}}^{z}=\sigma_{1}, \cdots
S_{x_{r}}^{z}= \sigma_{r})$ the probability that the spins at position 
$x_1, x_2 \cdots, x_r$ have a configuration $\sigma_{1}, \sigma_{2}, 
\cdots, \sigma_{r}$, with $\sigma_j=\pm 1$ for $j=1,2, \cdots, r$.
Then 
\begin{eqnarray}\label{wow}
P(S_{x_{1}}^{z}=\sigma_{1}, &\cdots& S_{x_{r}}^{z}=\sigma_{r})=
\frac{1}{Z(n,m)} \sum_{j_{1}=x_{1}-m}^{n}
F(j_{1})~q^{2x_{1}(1-\bar{\alpha}_1)} \;\;\;\;\;\;\; \nonumber\\ 
& & \sum_{j_{2}=j_{1}}^{\Delta x_{2}+t_{1}} \cdots 
\sum_{j_{r-1}=j_{r-2}}^{\Delta x_{r-1}+t_{r-2}}~\sum_{j_{r}=j_{r-1}}^{n}~  
\left (\prod_{k=2}^{r} F(j_{k-1},j_{k})~q^{2 x_{k}(1-\bar{\alpha}_{k})}\right ) 
F(j_{r}) \nonumber\\
\end{eqnarray}
where have simplified the notation for the partition functions by denoting 
\be 
F(j_{1})=Z(j_{1}-1,x_{1}-j_{1}), \;\;\; {\rm and} \;\;\; 
F(j_{r})=Z(j_{r}-\bar{\alpha}_{r},x_{r}-j_{r}+\bar{\alpha}_{r};n,m),
\ee
and 
\be\label{wow1}
F(j_{k-1},j_{k})=Z(j_{k-1}-\bar{\alpha}_{k-1},x_{k-1}-j_{k-1}+
\bar{\alpha}_{k-1}; j_{k}-1,x_{k}-j_{k}).
\ee
We have also introduced the new variables $\bar{\alpha}_{k}=1-\alpha_k$, 
$t_{k}=j_{k}-\bar{\alpha}_{k}$ and $\Delta=x_{k}-x_{k-1}$. 
\newline
The main result of this section is the following.

\begin{thm}[Exponential bounds on the correlations]\label{thm:exp_bound}
Let $v=\sum_{j=1}^{r}\alpha_j$ be the number of down spin on
the observable set $\alpha_1,\cdots, \alpha_r$, and let 
$d_k=(x_{k}-n)\alpha_k$ be the distances
of a down spin from the point of coordinate $n$ (distance to
the interface). Then the following bound holds
\be\label{UV}
P(S_{x_{1}}^{z}=\sigma_{1}, \cdots S_{x_{r}}^{z}=\sigma_{r}) 
\leq q^{v(v-1)+2\sum_{k=1}^{r}d_k} \; .
\ee
This result extends (and reduce to in the case r=2 with up and down
spin) the \eq{+-}. 
\end{thm}

\proof We now use the translation property to first shift partition function 
in the last box  
$B(j_{r}-\bar{\alpha}_{r},x_{r}-j_{r}+\bar{\alpha}_{r};n,m)$ in (\ref{wow1}) 
one unit to left (when $\bar{\alpha}_{k}=0$) or down (when
$\bar{\alpha}_{k}=1$). We get
\be\label{wow2}
Z(j_{r}-\bar{\alpha}_{r},x_{r}-j_{r}+\bar{\alpha}_{r};n,m)=
q^{2(n-j_{r}+\bar{\alpha}_{r})}~ Z(j_{r}-1,x_{r}-1;n-(1-\bar{\alpha}_{r}),
m-\bar{\alpha}_{r})
\ee
Next we estimate the factor of {\it q} above by one before 
substituting (\ref{wow2}) in (\ref{wow}), and we also factorize 
out all the bond weights $2x_{k}(1-\bar{\alpha}_{k})$, thus obtaining 
the bound
\begin{eqnarray}\label{wow3}
P(S_{x_{1}}^{z}=\sigma_{1}, &\cdots& S_{x_{r}}^{z}=\sigma_{r}) 
\leq q^{2\sum_{k=1}^{r}x_{k}(1-\bar{\alpha}_{k})}~ \frac{1}{Z(n,m)}
\sum_{j_{1}=x_{1}-m}^{n} Z(j_{1}-1,x_{1}-j_{1}) \nonumber\\ 
& & \sum_{j_{2}=j_{1}}^{\Delta x_{2}+t_{1}} \cdots 
\sum_{j_{r-1}=j_{r-2}}^{\Delta x_{r-1}+t_{r-2}}~ \left (
\prod_{k=2}^{r-1} F(j_{k-1},j_{k})~ \right)
Z_{r-1}(n-(1-\bar{\alpha}_{r}),m-\bar{\alpha}_{r}) \nonumber\\
\end{eqnarray}
where we have observed that by carrying out the summation 
over $j_{r}$ we obtain the partition function
\begin{eqnarray}\label{wow4}
Z_{r-1}(n-(1-\bar{\alpha}_{r}),m-\bar{\alpha}_{r})&=&\sum_{j_{r}=j_{r-1}}^{n} 
Z(j_{r-1}-\bar{\alpha}_{r-1},x_{r-1}-j_{r-1}+\bar{\alpha}_{r-1}; j_{r}-1,x_{r}-j_{r})
\nonumber \\
& & \;\;\;\; Z(j_{r}-1,x_{r}-1;n-(1-\bar{\alpha}_{r}),m-\bar{\alpha}_{r}).
\end{eqnarray}
From here we will need to repeat this procedure of performing shifts of
one unit down or to the left in succession to the partition functions in
(\ref{wow3}). After each step the resulting partition function obtained 
is changed according to the number of shifts we have performed.
At the end, we get  
\be\label{wow5}
P(S_{x_{1}}^{z}=\sigma_{1}, \cdots S_{x_{r}}^{z}=\sigma_{r}) 
\leq q^{2\sum_{k=1}^{r}x_{k}\alpha_{k}}~ 
\frac{Z(n-\sum_{k=1}^{r}\alpha_{k},~ m-\sum_{k=1}^{r} 
(1-\alpha_{k}))} {Z(n,m)}.
\ee
Substituting (\ref{wow6}) in (\ref{wow5}), with
$v=\sum_{k=1}^{r} \alpha_{k}$ yields the theorem.
\newline
\blackbox
\newline

Theorem \ref{thm:exp_bound} can be used to study the fluctuations around
the interface. In combination with the conservation of the third component
of the spin, the bound implies that fluctuations are strongly correlated.
In order to illustrate this we consider the total third component of the spin
on an interval centered on the interface: let $N$ and $L$ be even positive
numbers, $L\leq N$, and consider the state $\psi(N/2,N/2)$, and let $\langle
\cdot\rangle_N$ denote the expectation in this state. Define $F_L$ by
\be
F_L=\sum_{x=(N-L)/2+1}^{(N+L)/2} S^z_x
\ee
We will also need the total third component of the spin in the complement
of the interval $[(N-L)/2+1,(N+L)/2]$, defined by
$$
F_L^c=\sum_{x\not\in[(N-L)/2+1,(N+L)/2]} S^z_x
$$
Then, for all $L\leq N$, 
$$\langle F_L\rangle_N=0$$ 
as a consequence of the symmetry properties given in Theorem \ref{symm}.

\begin{thm}\label{thm:lim_dist}
\be
\lim_{L\to\infty} \lim_{N\to\infty}Prob_N(F_L=l)=\delta_{l,0}
\ee
\end{thm}
\proof
The distribution of $F_L$, and, hence, its variance, can be estimated 
by first noting that 
$$
Prob_N(F_L=l)=Prob_N(F^c_L=-l)=Prob_N(F^c_L=l)
$$
and further that
$$
Prob_N(F^c_L=l) \leq Prob_N(\mbox{there are at least $l$ down spins
in $[(N+L)/2+1,N]$})
$$
By summing the bound of Theorem \ref{thm:exp_bound} over all numbers 
$r\geq l$ of down spins to the right of $(N+L)/2$, and possible positions 
$x_1,\ldots,x_r$, we obtain the following bound:
\beann
\lefteqn{Prob_N(\mbox{there are at least $l$ down spins
in $[(N+L)/2+1,N]$})} \;\;\;\;\;\;\;\;\;\;\;\;\;\;\;\;\;\;\;\;
\;\;\;\;\;\\
&\leq&
\sum_{r=l}^{(N-L)/2}\sum_{(N+L)/2<x_1<\cdots<x_r\leq N}
q^{r(r-1)+2\sum_{k=1}^r (x_k-N/2)} \;\;\;\;\;\;\;\;\;\;\;\;\\
&\leq& \sum_{r=l}^\infty \frac{q^{r(r-1)}}{r!}\left[\sum_{x=L/2+1}^\infty
q^{2x}\right]^r\\
&\leq& \sum_{r=l}^\infty \frac{q^{r(r-1)}}{r!}\left[\frac{q^{L+2}}{1-q^2}
\right]^r\\
&\leq& q^{l(l-1)} \frac{1}{l!}\left[\frac{q^{L+1}}{1-q^2}\right]^l
\exp[q^{L+3}/(1-q^2)]\leq C(q) q^{l^2+Ll}
\eeann
where $C(q)$ is a constant depending only on $q$. From this bound it is clear that
\be
\lim_{L\to\infty} \limsup_{N\to\infty}Prob_N(F_L=l)=\delta_{l,0}
\ee
As it has been shown in \cite{GW} that the limit $N\to\infty$ 
exists, this concludes the proof.
\blackbox

\masect{Higher Dimensions}
\noindent
The path integral formulation we have introduced provides an efficient 
way to bound correlation functions of the quantum XXZ model in one
dimension. For higher dimensions it is known that the state with {\it n}
spins down in {\it d} dimensions is given by \cite{ASW}
\be
\psi(n,m)=\sum_{\alpha_{x}} \left \{ \prod_{x}
q^{\alpha_{x}|x|} \right \} \mid \{ \alpha_{x} \} \rangle.
\ee
where $|x|$ is the $L_1$ norm of the vector $x$. 
\newline
We consider a two-dimensional spin system in order to illustrate 
how to relate the property of the model in higher dimensions to those
of a one dimensional system. 
\newline
Since we are free to choose the orientation of the physical spin system, 
we prefer to dispose the spins along {\it M} diagonal lines with each
diagonal having the same number {\it N} of spins, as in the first
diagram of Fig. 9. The weights assigned to the bonds in the corresponding
path representation follow the diagonal pattern shown in the second
diagram of Fig. 9. The analytic expression for the weight of a bond in
this case is  
\be\label{bond4}
w(b)=\left \{
\begin{array}{ll}
x+y\;\;\; {\rm any~ horizontal~ bond~ ending~ at}~ (x,y) \\
1 \;\;\;\;\;\;\;\;\;\;\; {\rm any~ vertical~ bond} 
\end{array}
\right.
\ee
\begin{figure}
\centerline{
\epsfbox{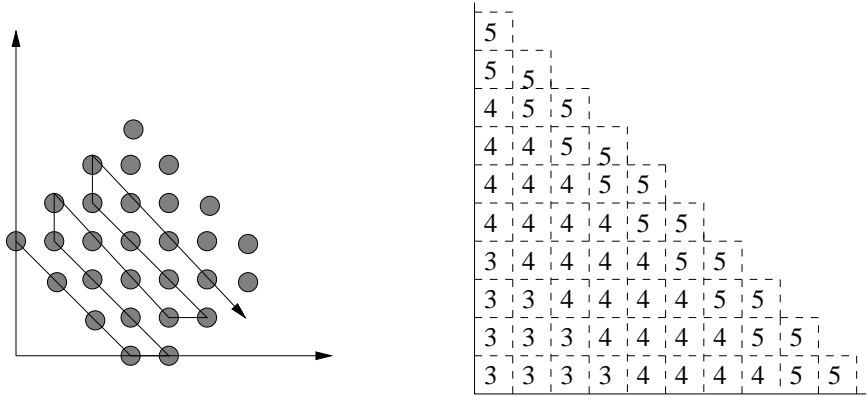}}
\caption{{\sl The spin system in any dimension can be put in
correspondence with a one-dimensional spins system by using the Cantor
diagonal procedure as indicated in the diagram for the two-dimensional
case.}}
\end{figure}
Our result is based on \eq{TF} and shows in detail the mechanism
of dimensional reduction which underlines the methods used to prove to 
absence of gap for interface excitations in $d=3$ \cite{BCNS}.
\newline
The main result of this section is the following theorem.
\begin{thm}
Consider a two-dimensional system shown in Fig.10a, having sizes $N$
and $M$. Let ${\cal K}$ be the set of $m$ non-negative integers $\{k_i\}$
such that $\sum_{i=0}^{m}ik_i=k$ and $\sum_{i=0}^{m}k_i=N$. The norm of
the ground state of the two-dimensional system with $k$ spins down,
$Z_{2d}(k,NM-k)$, is given by
\be
Z_{2d}(k,NM-k)=q^{2(N-1)k} \sum_{\{k_i\}\in {\cal K}}
\frac{N!}{k_{0}!k_{1}! ... k_{m}!} \prod_{i=1}^{m} 
\left \{ Z(j,M-j) \right \}^{k_{j}} \; .
\label{wkh}
\ee
where $Z(j,M-j)$ are the partition functions of the one-dimensional
model.
\end{thm}
{\bf Examples.} To illustrate the theorem let us calculate the 
two-dimensional partition function for a system of 9 spins with $N=3,
M=3$. We take $k=3$. In this case, there are three set of allowed values
of $(k_{0},k_{1},k_{2},k_{3})$. They are $(2, 0, 0, 1)$, $(1, 1, 1, 0)$
and $(0, 3, 0, 0)$. Thus we obtain  
\be
Z_{2d}(3,6)=q^{12} \left \{ {Z(1,2)}^{3}+6Z(1,2)Z(2,1)+3Z(3,0) \right \}
\ee
When $n=4, m=5$, the following are the sets of allowed {\it k} values
are $(2, 0, 0, 1)$, $(1, 1, 1, 0)$ and $(0, 3, 0, 0)$. We get
\be
Z_{2d}(4,5)=q^{16} \left \{ 6Z(1,2)Z(3,0)+3{Z(2,1)}^{2}+3Z(3,0) \right \}
\ee
\proof Because of the periodic pattern of the weights in path space, 
the grand-canonical partition function in two-dimensions is given by 
\be\label{GC1}
Z_{GC}^{2d}=\prod_{j=N}^{N+M-1}
(1+zq^{2j})^{N}=\sum_{k=0}^{NM}z^{k}Z_{2d}(k,NM-k),
\ee
where $Z_{2d}(k,NM-k)$ is the canonical partition function in
two-dimensions.
\newline 
The product formula in (\ref{GC1}) can also be written in terms 
of the generalized canonical partition functions of the one-dimensional
system. By interchanging the $N$-th power with the product in \eq{GC1}
we get
\be\label{GC3}
\prod_{j=N}^{N+M-1} (1+zq^{2j})^{N}=\left \{ \sum_{l=0}^{M} z^{l}
Z_{1d}(N-1,0; N-1+l,M-l) \right \}^{N}
\ee 
where the generalized partition functions have initial points
$(N-1,0)$ in order to account of the proper relation between the
weights of the corresponding one-dimensional and two-dimensional 
systems as we have defined them.
\newline
Now we use the translation property (\ref{TF}) to shift the generalized
partition functions in (\ref{GC3}) to the origin. In doing this we obtain 
a multiplicative factor depending on the first set of weights of the
two-dimensional system:
\be
Z_{1d}(N-1,0; N-1+l,M-l)=q^{2(N-1)l}Z_{1d}(l,M-l)
\ee
Substituting the above expression into (\ref{GC3}) we get
\be\label{GC4}
\prod_{j=N}^{N+M-1} (1+zq^{2j})^{N}=\left \{ \sum_{l=0}^{M} z^{l} 
q^{2(N-1)l}Z_{1d}(l,M-l) \right \}^{N}.
\ee
Equating (\ref{GC4}) to the expression on the right hand side
of (\ref{GC1}) yields
\be\label{equal}
\sum_{k=0}^{NM}z^{k}Z_{2d}(k,NM-k) =\left \{ \sum_{l=0}^{M} z^{l} 
q^{2(N-1)l}Z_{1d}(l,M-l) \right \}^{N}
\ee
Since the equality in (\ref{equal}) holds term by term in powers of 
{\it z}, we express the two-dimensional partition function as a sum of
one-dimensional partition functions given by
\be
Z_{2d}(k,NM-k)=q^{2(N-1)k} \sum_{k_{0}, k_{1},k_{2}, ... k_{m}}
\frac{N!}{k_{0}!k_{1}! ... k_{m}!} \prod_{i=1}^{m} 
\left \{ Z_{1d}(j,M-j) \right \}^{k_{j}}
\ee
where the sum runs over the values of {\it k} with the restrictions
$k_{1}+2k_{2}+3k_{3}+...+mk_{m}=k$, and $k_{0}=N-\sum_{i=1}^{m} k_{i}$.
\newline
\blackbox
\newline

\noindent {\large \bf Acknowledgments\/}
P.C. thanks G. Kuperberg for several interesting discussions 
on counting and q-counting problems. O.B. was supported by 
FAPESP under grant 97/14430-2. B.N. acknowledges partial 
support by NSF under grant DMS-9706599.

\end{document}